\documentclass[review,1p,number,sort&compress]{elsarticle}
\usepackage{amssymb}
\usepackage{lipsum}
\usepackage{siunitx}
\usepackage{pdflscape}
\usepackage{mhchem}
\usepackage{url}
\usepackage{comment}
\usepackage{caption}
\usepackage{subcaption}
\usepackage[normalem]{ulem}
\usepackage[utf8]{inputenc}
\usepackage[T1]{fontenc}
\usepackage{overpic}

\journal{Nuclear Instrumentation and Methods A}

\begin{document}

\begin{frontmatter}



%\title{Photon-to-Digital Converters for Particle Physics: a User Perspective}
\title{Proof-of-Concept and User Perspective on Photon-to-Digital Converter Applications in Particle Physics}


\author[1]{M. \'A. Garc\'ia-Peris\corref{corr}}
\ead{miguel.garciaperis@manchester.ac.uk}
\author[1]{B. Palmeiro}
\author[2]{G. Lessard}
\author[1]{R. Guenette}
\author[2]{S.A. Charlebois}
\author[1]{E. Gramellini}
\author[2]{J.-F. Pratte}
\author[2]{T. Rossignol}
\author[2]{N.Roy}
\author[2]{F. Vachon}

\affiliation[1]{Department of Physics and Astronomy, University of Manchester, United Kingdom}
\affiliation[2]{Interdisciplinary Institute for Technological Innovation and Department of Electrical and Computer
Engineering, University of Sherbrooke, Canada}

\cortext[corr]{Corresponding author}
\date{\today}

\begin{abstract}
Photon-to-Digital Converters (PDCs) are photosensors with single photon resolution and large dynamic range that digitise the output of individual single-photon avalanche diodes directly on-chip, offering potential advantages with respect to analogue Silicon Photo-Multipliers (SiPMs). In this work, we assess the PDC technology from the perspective of particle physics, using early, low-coverage PDC prototypes for proof-of-concept studies in calorimetry and tracking. Under simple, but extrapolable, detector conditions, we qualitatively compare PDCs with SiPM-based systems to illustrate their potential in terms of detector performance and readout. As the first application of this technology to particle physics instrumentation, we place particular emphasis on the user perspective, reporting on the practical experience of operating PDCs — from setup and calibration to data handling — and highlighting the relative ease of integration compared to conventional SiPM readout. We discuss the potential of PDCs to address key challenges in large-scale instrumentation, and their prospective integration into next-generation high-energy physics experiments.
%Photon-to-Digital Converters (PDCs) are photosensors with single photon resolution and large dynamic range that digitise the output of individual single-photon avalanche diodes directly on-chip, offering potential advantages with respect to analogue Silicon Photo-Multipliers. In this work, we asses the PDC technology from the perspective of particle physics, using low-coverage PDC demonstrators for proof-of-concept studies in calorimetry and tracking. Under simple, but extrapolable, detector conditions, we qualitatively evaluate the potential of PDCs to improve detector performance and readout compared to SiPM-based systems, and discuss their potential to address key challenges in large-scale instrumentation and its potential integration into next-generation high-energy physics experiments.
\end{abstract}



\begin{keyword}
%% keywords here, in the form: keyword \sep keyword, up to a maximum of 6 keywords
Photon-to-Digital Converters (PDCs) \sep Digital SiPMs \sep Light detection \sep Calorimetry \sep Time-correlated single photon counting



\end{keyword}

\end{frontmatter}

%\tableofcontents


\section{Introduction}

Light detection with single-photon resolution is fundamental in high energy physics. The Photo-Multiplier Tube (PMT) technology~\cite{bib:pmts} was among the first to provide such an achievement, and after several decades of development remains a well-understood and integrated approach with a proven track record in multiple particle physics experiments~\cite{bib:sk_tdr,bib:juno_tdr,bib:minos_tdr,bib:icecube_tdr,bib:borexino_tdr}. %PMTs consists of a photoelectric or photosensitive material that converts photons into electrons, which are further amplified by a succession of dynodes to generate a measurable current. Available in different sizes, geometries and wavelength-sensitiveness, they can be adapted to different detectors concepts and coverages, fulfilling its physics requirements.

More recently, Silicon Photo-Multipliers (SiPMs) have been widely adopted for many applications, especially in particle and nuclear physics instrumentation, being commonly used across many experiments~\cite{bib:SIMON201985, bib:Chepel_2013, bib:11287941}. They consist of an array of Single-Photon Avalanche Diodes (SPADs), with sizes ranging  \qtyrange{10}{100}{\micro\meter}, for a typical active area of \qtyrange{1}{100}{\square\milli\meter}~\cite{bib:gundacker,bib:piemonte}. Compared with PMTs, which are bulkier and require operation voltages of \qtyrange{0.1}{1}{\kilo\volt}, SiPMs are easily adaptable to different particle physics experiments due to their lower operating voltage (\qty{\sim 50}{V}), compact design and insensitivity to magnetic fields, while offering similar photodetection efficiency, gain, and response time. However, SiPMs present other challenges. The analogue nature of the SiPM output, characterised by small single-photoelectron signals and a wide dynamic range, requires a carefully designed readout chain~\cite{bib:gundacker,bib:piemonte}. In addition, their intrinsic noise is significantly higher than that of PMTs, with uncorrelated dark count rate reaching $\mathcal{O}$(\qtyrange{0.1}{1.0}{\mega\hertz/\milli\meter^{2}}), and the correlated optical crosstalk and afterpulsing with typical values of $\mathcal{O}$(\qtyrange{1}{10}{\%})~\cite{bib:DU2008396,bib:Wang:2025pxy,bib:Para:2015rfa}. Finally, their smaller active area requires of a larger number of channels for the same optical coverage.

From a scalability perspective, the readout of a large number of SiPMs further increases the system complexity. Operating large arrays of SiPMs simultaneously leads to a substantial increase in the number of readout channels and in the overall data throughput, placing stringent requirements on the front-end electronics and data acquisition system~\cite{bib:ganging_1}. Although sensor ganging is commonly used to mitigate channel count, with different combinations of serial and parallel~\cite{bib:ganging_2}, it introduces a trade-off by increasing the effective input capacitance per channel, which degrades signal bandwidth and timing performance. As a reference for the scale of such systems, large SiPM-based detector implementations typically involve hierarchical readout architectures with on the order of $10^{3}$–$10^{4}$ electronic channels instrumenting $10^{5}$–$10^{6}$ photosensors~\cite{bib:dune_tdr4,bib:ds_tdr,bib:nexo_tdr}, depending on the level of aggregation adopted in the front-end design. This illustrates the inherent tension between readout simplification and preservation of signal fidelity and granularity in high-channel-count detector systems.

These limitations motivated the development of alternative readout concepts aimed at preserving the intrinsic digital information of SPADs while reducing overall system complexity~\cite{bib:mit_1998}. Photon-to-Digital converters (PDCs), or more generally, digital SiPMs, have been proposed as a next step in photon sensor technology~\cite{bib:pdc_pratte}. The central idea is to exploit the intrinsic boolean response of individual SPAD cells. In conventional SiPMs, current signals from all cells are summed in parallel, resulting in the loss of this binary information and producing an analogue output. In contrast, the PDC technology provides a direct readout of individual cells, thereby retaining their digital nature.

In principle, and depending on the specific implementation, this approach can address several of the limitations associated with analogue SiPMs~\cite{bib:charbon_pdc_1,bib:charbon_pdc_2,bib:gundacker_pdc_comparison}. Since the signal remains digital throughout the readout chain, SPAD-to-SPAD gain variations are mitigated and the number of fired cells can be obtained directly from the digital output. This avoids analogue amplification, signal calibration and Analogue-to-Digital Converter (ADC) resolution optimization. At the system level, it opens up new opportunities for signal processing and data acquisition, such as sparse or event-driven operation modes, potentially reducing data throughput and storage, and power consumption. Noisy SPADs can be selectively disabled and SPAD hold-off can be artificially increased to reduce the impact of dark count rate and after pulsing, respectively, allowing the user to find an optimal trade-off between noise, effective sensitive area, and detection time. Although the magnitude of these gains depends strongly on the chosen implementation and readout architecture~\cite{bib:henderson_1}, its potential has been studied in fields such as medical imaging~\cite{bib:pet,bib:gundacker_pdc_comparison}, imaging and adaptative optics~\cite{bib:canon}, and military~\cite{bib:mit_2016}.  

%Although first applications of digital SiPMs are beginning to emerge in particle-physics contexts~\cite{bib:desy_testbeam,bib:platon,bib:tracking}, this technology remains relatively unfamiliar to the broader HEP community, and a clear path towards its wider adoption has yet to be established. One possible reason is the lack of an evaluation from the detector-user perspective that translates the potential benefits at the electronic level into tangible benefits for particle-physics applications. This work aims to fill this gap by presenting a proof-of-concept study based on simple use cases to assess the feasibility of operating such devices within the context of particle physics. 
The use of digital SiPMs in particle physics is a very recent development, with only a handful of early studies to date~\cite{bib:desy_testbeam,bib:platon,bib:tracking,bib:Fischer:2025gqn}, and the technology remains relatively unfamiliar to the broader high-energy-physics community. In this work, we present the first evaluation of PDCs from the perspective of a potential user, employing early PDC prototypes in a proof-of-concept study built on simple test cases to assess both the feasibility and the potential of operating such devices within the context of particle physics. In particular, we qualitatively compare their behaviour with that of conventional analogue SiPMs, in order to provide a basic understanding of the PDCs' signal response and operational characteristics. Rather than providing a full performance optimisation and exhaustive comparison with conventional technologies, the goal of this study is first, to address the ease of use of this novel technology by a researcher familiar with analogue SiPMs or PMTs, and second, to identify key operational features that could be of benefit for future detector applications. The device's performance and test cases are therefore discussed from this broader perspective.

\section{PDC technology and experimental set-up}\label{section:PDC}
PDCs are an emerging class of digital SiPMs and the details of their technology can be found in multiple references~\cite{bib:Rossignol:2024ylw,bib:sherbrooke_1,bib:sherbrooke_2, bib:s23073376, bib:s21020598,10414786}. They consist of an array of SPADs, each individually connected to a dedicated quenching circuit and read out independently. They produce a binary signal upon the occurrence of an avalanche, after which they are quenched and remain inactive for a configurable hold-off time. In this way, the avalanche detection is intrinsically discrete with no analogue electronic noise added, and the information can be processed in digital form throughout the entire readout chain. 

%The digital architecture also enables the assignment of event-level flags, encoding additional information such as temporal or multiplicity conditions. In this work, these flags are used for event selection and noise suppression.

Different technological implementations of digital SiPMs have been developed~\cite{bib:henderson_1}, primarily distinguished by the integration of the sensor and readout electronics. In 2D architectures, both the SPAD array and the electronics are implemented on the same substrate~\cite{bib:Fischer:2025gqn,bib:philips_1,bib:philips_2}, whereas in 3D integration these layers are stacked, enabling a larger fill factor~\cite{bib:henderson_1,bib:mit_2016,bib:canon,bib:stmicro,bib:sony}. The device used in this work is an early 2D implementation on the 3D architecture developed by the University of Sherbrooke and fabricated by Teledyne~\cite{bib:Rossignol:2024ylw,bib:sherbrooke_1,bib:sherbrooke_2, bib:s23073376, bib:s21020598,10414786}. 

\begin{figure}[htbp]
\centering
\begin{overpic}[width=0.9\textwidth]{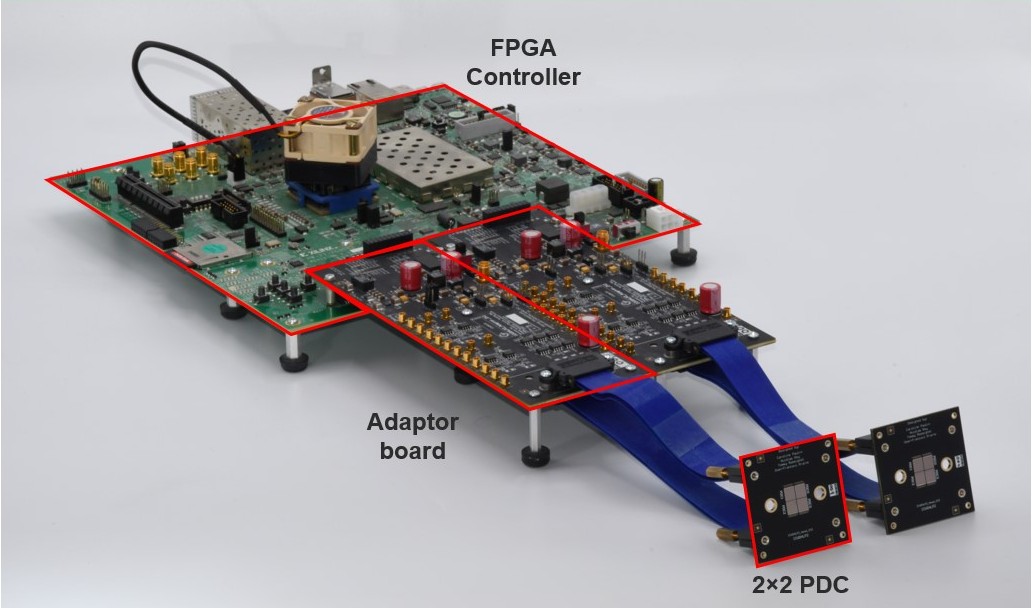}
  \put(0,0){\includegraphics[width=0.3\textwidth]{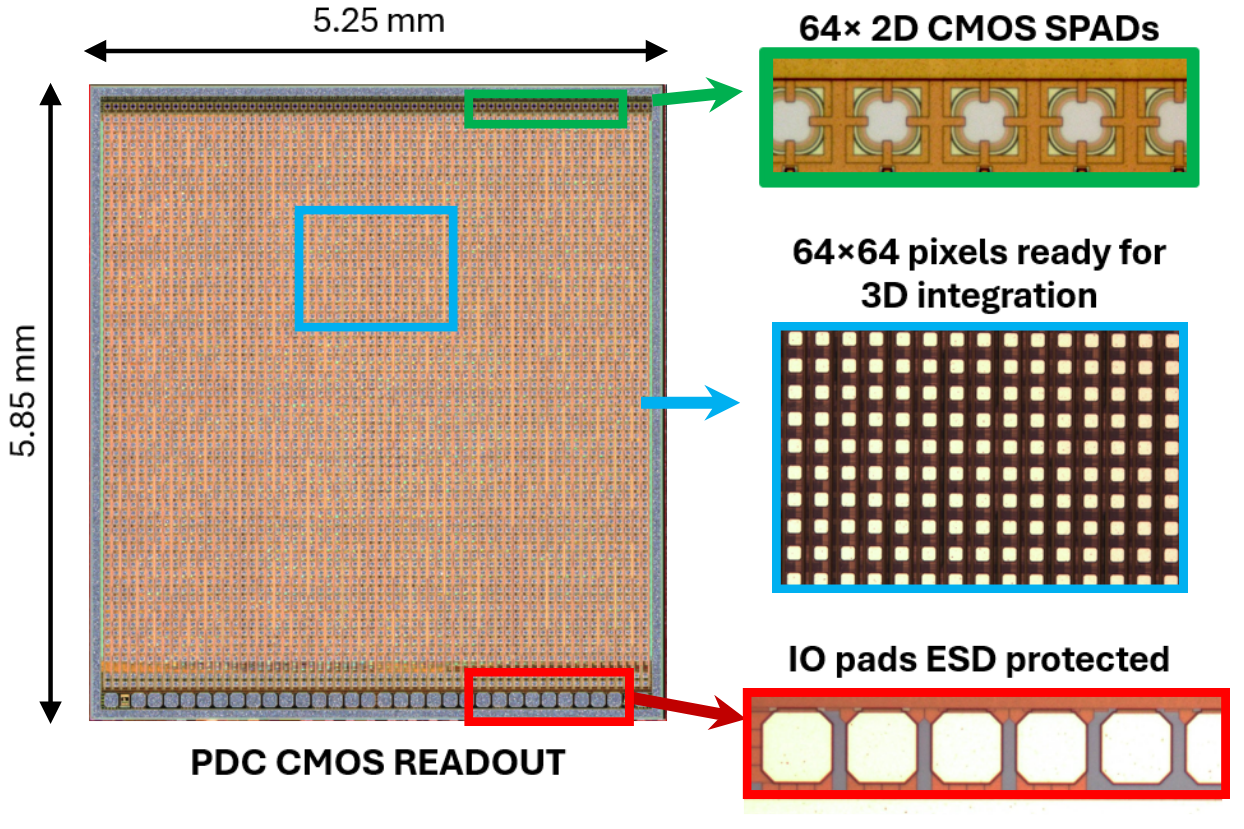}}
\end{overpic}
\caption{PDC prototypes and custom readout system used in this study, with the different parts labelled and framed in red: the FPGA, custom-made adaptor boards and PDC boards. In this case, two PDC PCBs are connected in the system. The top-left image illustrate the prototype layout, where only 64 SPADS are active.
\label{fig:fpga}}
\end{figure}

The system employed in this study is presented in Fig.~\ref{fig:fpga}, corresponding to an early implementation of the PDC concept for preliminary testing, debugging and FPGA programming. It consists of printed circuit boards (PCBs) hosting four PDC prototypes, each comprising only 64 2D SPADs of \qty{21}{\micro\meter} of side integrated in the Complementary Metal-Oxide-Semiconductor (CMOS) readout. The devices are interfaced via wire bonding and operated using a Field-Programmable Gate Array (FPGA) based readout system, which configures, controls and processes the digital response of the detector under different experimental conditions. The FPGA used is a ZCU102 model from Xilinx~\cite{bib:fpga} and the adaptor boards are custom-made. The system records the number of fired SPADs as a function of time, with a minimum time-bin of 10 ns. More advanced versions of the system (now available, yet not at the time of this work) present the complete 3D architecture with a matrix of 64$\times$64 SPADs~\cite{bib:3d_last}, and are expected to reach a time binning of the order of hundreds of picoseconds.

\section{Experimental tests}
This section presents a set of proof-of-concept measurements designed to explore the operation of PDCs from the perspective of a researcher familiar with conventional SiPM- or PMT-based detectors. Rather than providing a comprehensive characterisation of the devices, the measurements focus on simple and representative test cases relevant to particle and nuclear physics. The PDCs were first exposed to a controlled light source, provided by a light-emitting diode (LED), to investigate their basic signal response and operational characteristics. They were subsequently used to characterise the spectrum of a radioactive source and to detect the passage of cosmic rays, illustrating their response to increasingly realistic particle-physics signals. The first two measurements were performed alongside a conventional analogue SiPM, allowing a qualitative comparison of the two technologies and providing a familiar reference for interpreting the PDC response.

\subsection{Basic performance of PDCs}

For the first test case, the four PDC prototypes described in section~\ref{section:PDC} are evaluated using commercial LEDs peaked at 470 nm as light sources, and compared with analogue SiPMs for a qualitative understanding of their benefits. The analogue SiPM used was a Hammamatsu S13370-3050CN~\cite{bib:sipm_hpk} operated at 53.5 V, the signal of which was amplified with a custom-made amplification board based on the TI LMH6624 amplifier~\cite{bib:amplifier}, and finally read out by a Teledyne Lecroy WaveSurfer 4054HD oscilloscope. All devices were placed in a sealed dark box to avoid background light.

For the initial step, both devices were exposed to the LED operated under various driving conditions to produce light pulses of varying intensity. Given the differences in active area and SPADs between the PDCs (4 PDCs with 64 SPADs of \qty{21}{\micro\meter} pixel size) and the SiPM (3600 SPADs of \qty{50}{\micro\meter} pixel size) the dynamic range is very different between both technologies. Therefore, the experimental configuration was kept identical in terms of geometry and illumination, placing both sensors at the same distance from the light source and operating the LED under the same voltage settings, such that the optical flux in both devices was consistent. No attempt was made to perform an absolute calibration of the photon flux; instead, the goal of this measurement is to provide a qualitative comparison of the basic signal response of both devices under nominally similar illumination conditions. The pulse generator used to fire the LED was also used to externally trigger the corresponding readout system. 

\begin{figure}[tbp]
\centering
\includegraphics[width=0.49\textwidth]{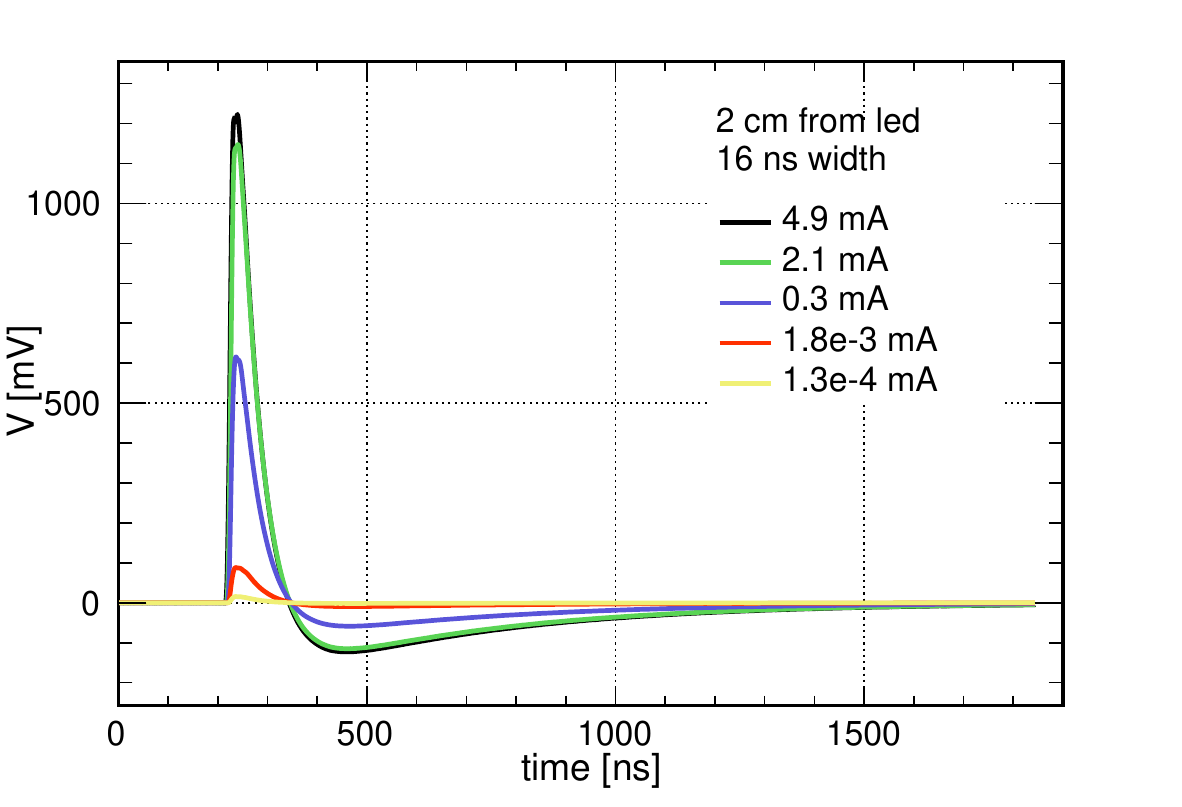}
\includegraphics[width=0.49\textwidth]{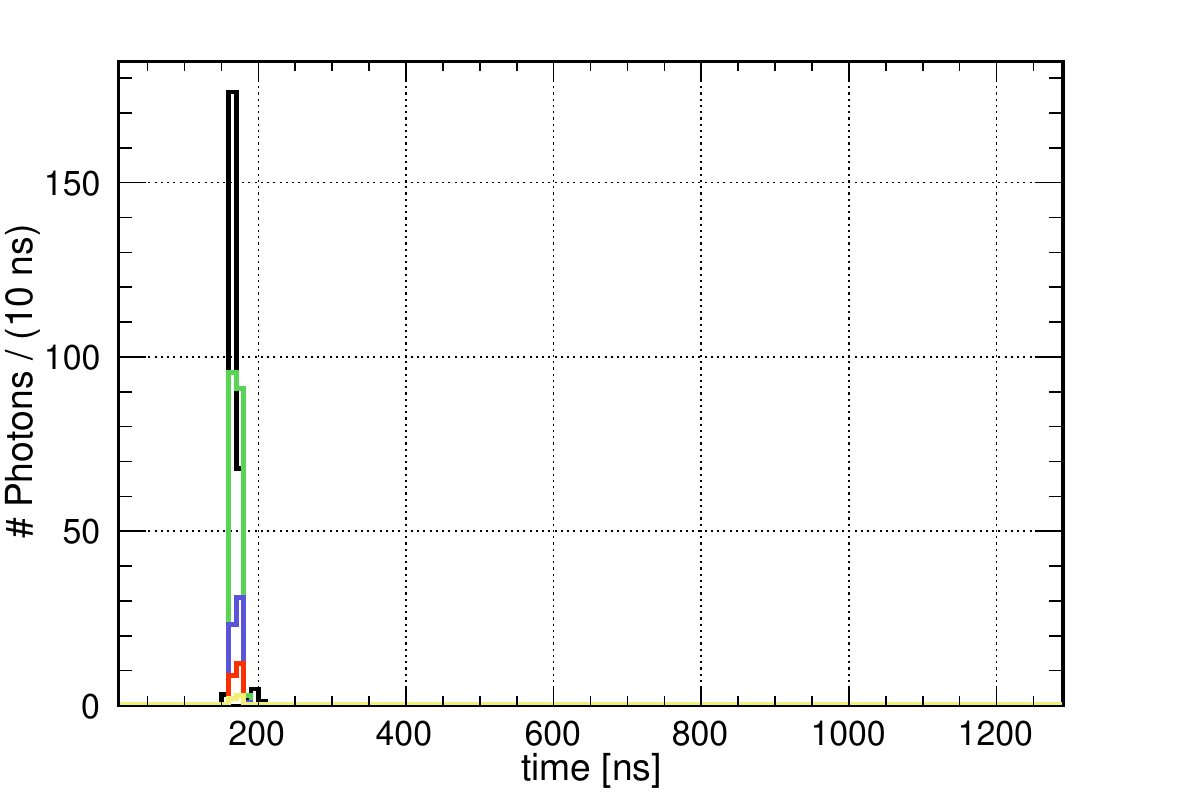}
\caption{Analogue (left) and PDCs output waveform to different intensity light pulses. Each waveform shows the average of at least 1000 waveforms. Each colour represents a different LED intensity. The pulse is a trapezoid of 16 ns with 8 ns of rise and fall time
\label{fig:wfs}}
\end{figure}

Fig.~\ref{fig:wfs}-left shows the analogue SiPM response to the different light pulses. These were trapezoids of 16 ns with 8 ns of rise and fall time\footnote{Even though the voltage shape is known, the high non-linearly of the LED does not allow to know the exact temporal light profile with precision.}. As can be observed, more intense light signals generate larger output amplitudes. The readout chain introduces an undershoot in the signal, which increases with the size of the signal. In contrast, Fig.~\ref{fig:wfs}-right shows the corresponding response of the PDCs under equivalent illumination conditions. In this case, the readout provides a direct time-resolved count of triggered SPADs, namely, the direct number of collected photons within a configurable time interval (10 ns in this case). The measured output, therefore, encodes the temporal evolution of detected photons in terms of discrete triggered events, avoiding the need for analogue charge integration and reconstruction procedures. The interpretation of the output is thereby straightforward, as it is not affected by the analogue signal shaping (rise and decay), amplification and digitization artefacts of the front-end electronics chain used to read the SiPM. 

\begin{figure}[tbp]
\centering
\includegraphics[width=0.49\textwidth]{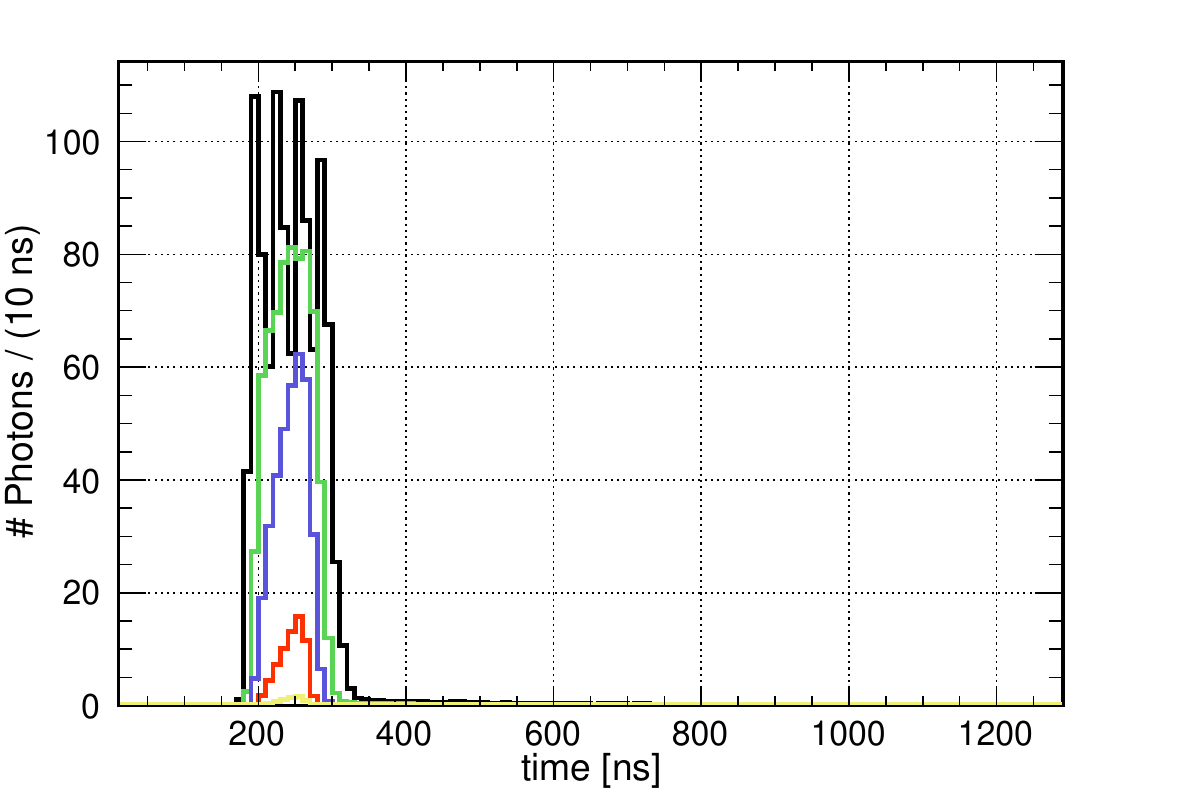}
\includegraphics[width=0.49\textwidth]{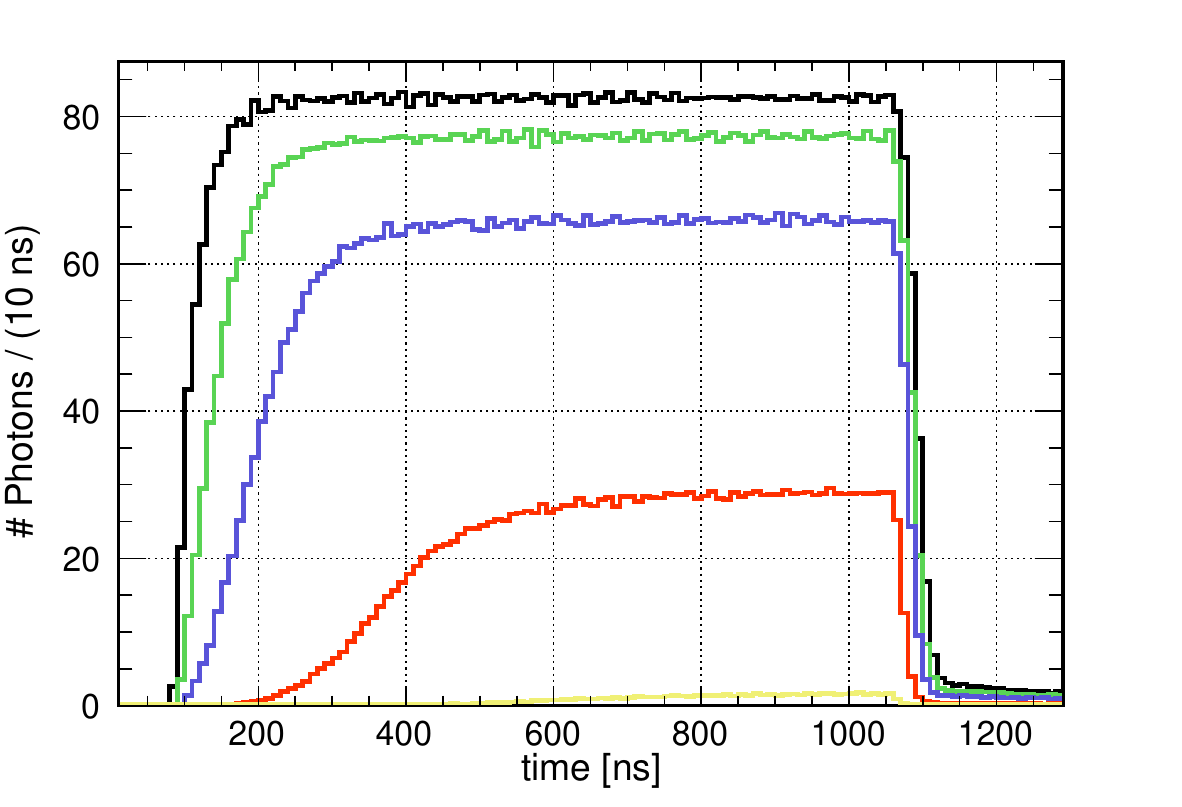}
\caption{PDCs output waveform to different trapezoidal-shaped light pulses. Left: 100 ns light pulse with 10 ns of rise and fall time. Right: 1000 ns light pulse with 10 ns rise and fall time. Each waveform shows the average of at least 1000 waveforms. Each colour represents a different LED intensity. 
\label{fig:wfs2}}
\end{figure}

Motivated by this simplicity, the PDCs were exposed to light pulses of longer duration and different intensities. The observed signals are presented in Fig.~\ref{fig:wfs2}. The pulses were in both cases trapezoids with 10 ns of rise and fall time, first of 100 ns wide (left) and later of 1000 ns wide (right). The temporal profile of the optical pulse, distorted by the high non-linearity of the LED, is readily inferred from the PDCs' response, as well as the relative amplitudes. Additionally, the effect of saturation is clearly visible in the measured signal. At high light intensities, most SPADs are triggered within a time bin, leaving fewer detectors available in the following bins due to the hold-off time. As the SPADs return to the ready state, the measured photon count increases again, producing a periodic modulation that can be corrected based on the number of SPADs triggered in the previous time bin. In an analogue SiPM, the signal interpretation would have been less direct. For pulse widths exceeding the recovery time of the SPADs ($\mathcal{O}(10 \mathrm{ns})$), individual cells can fire multiple times during a single light pulse. These successive avalanches are integrated into the analogue waveform and combined with the temporal response for the readout chain, including its undershoot, making the underlying temporal evolution of the signal less straightforward to reconstruct.

% \begin{figure}[tbp]
% \centering
% % ---------- LEFT COLUMN ----------
% \begin{minipage}[t]{0.47\textwidth}
% \centering
% \textbf{SiPMs}\par\vspace{0.5em}

% \includegraphics[width=\textwidth]{text/images/SiPMSignal_plot0.pdf}

% \vspace{0.5em}

% \includegraphics[width=\textwidth]{text/images/SiPMSignal_plot1.pdf}

% \vspace{0.5em}

% \includegraphics[width=\textwidth]{text/images/SiPMSignal_plot2.pdf}
% \end{minipage}
% \hfill
% % ---------- VERTICAL LINE ----------
% \vrule width 1pt
% \hfill
% % ---------- RIGHT COLUMN ----------
% \begin{minipage}[t]{0.47\textwidth}
% \centering
% \textbf{PDCs}\par\vspace{0.5em}

% \includegraphics[width=\textwidth]{text/images/PDCsSignal_plot0.pdf}

% \vspace{0.5em}

% \includegraphics[width=\textwidth]{text/images/PDCsSignal_plot1.pdf}

% \vspace{0.5em}

% \includegraphics[width=\textwidth]{text/images/PDCsSignal_plot2.pdf}
% \end{minipage}

% \caption{
% Analog (left) and digital (right) SiPM output waveform to different trapezoidal-shaped light pulses. Each waveform shows the average of at least 1000 waveforms. Each colour represents a different LED intensity. Top row: 16 ns light pulse with 8 ns of rise and fall time, and 2 cm separation from the LED. Mid row: 100 ns light pulse with 10 ns of rise and fall time, and 10 cm separation from the LED. Bottom row: 1000 ns light pulse with 10 ns rise and fall time, and 16 cm separation from the LED.
% }
% \label{fig:wfs}
% \end{figure}

\begin{figure}[tbp]
\centering
\includegraphics[width=0.49\textwidth]{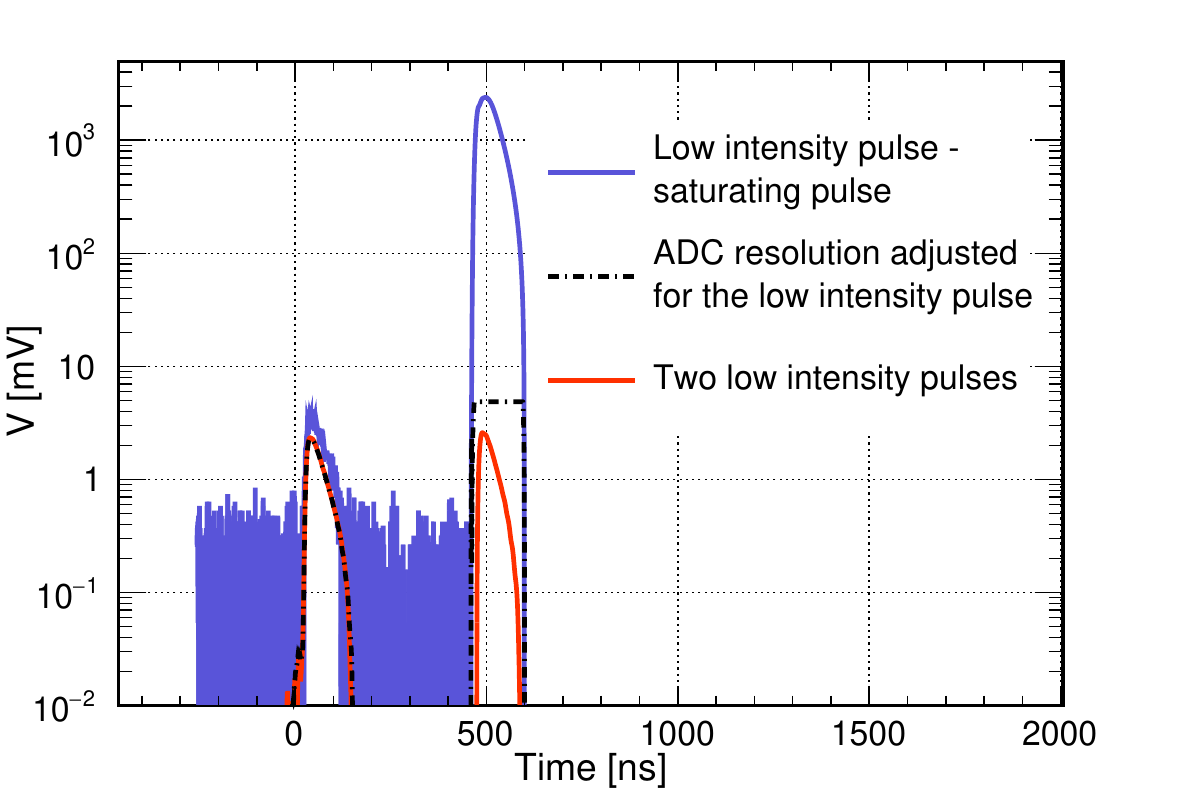}
\includegraphics[width=0.49\textwidth]{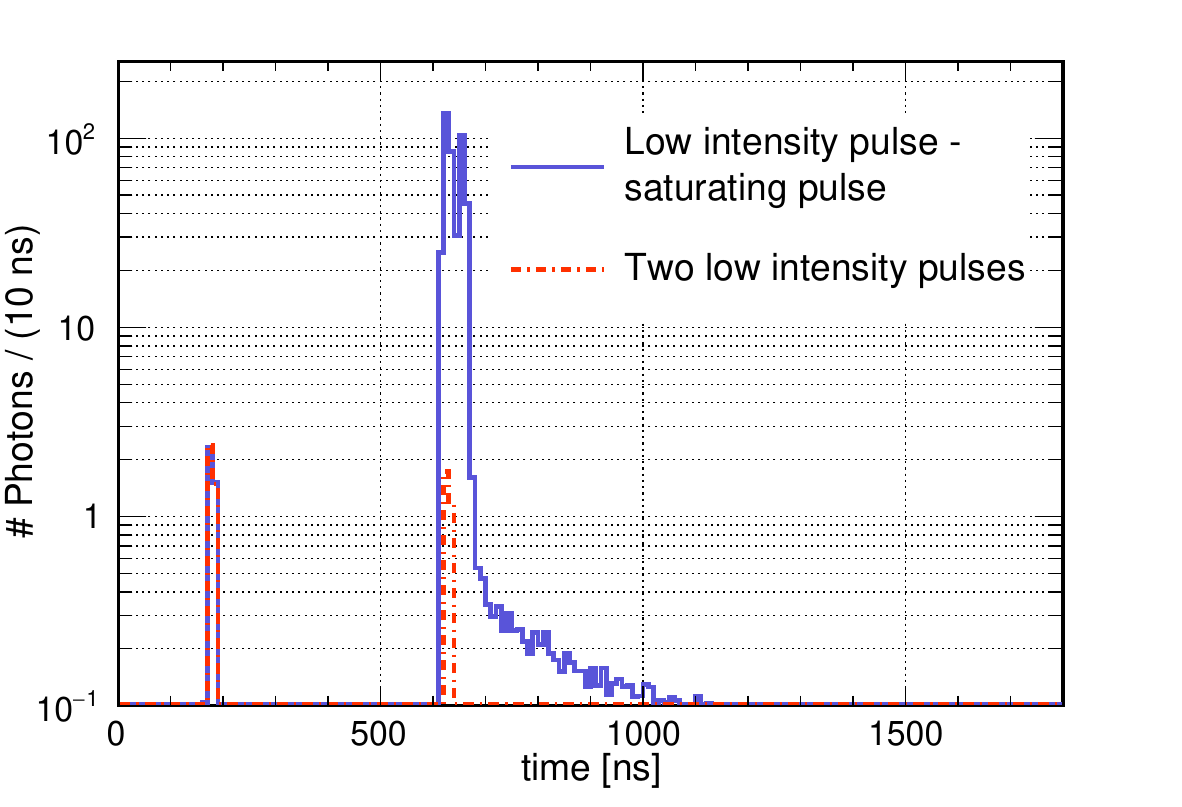}
\caption{SiPM response when exposed to two consecutive light pulses (left). The red line shows the case of two low-intensity pulses ($\mathcal{O}$(\qty{1}{photon})) with the oscilloscope resolution tuned to optimally capture this signal. The second light pulse is increased until the device is saturated (blue), and the oscilloscope resolution is optimised to capture both pulses. A degradation of the first pulse can be observed. Finally, the same pulse is captured, setting the oscilloscope resolution to the same value used in the original low-intensity configuration, thereby losing information about the second saturating peak (dashed black). In contrast, when the PDCs (right) are exposed to two low-intensity pulses (in dashed red), and then to a low-intensity pulse and a saturating pulse (solid blue), the first signal is observed without distortion in both cases. For both devices, the signals displayed are the averaged of 100 events.
\label{fig:adcres}}
\end{figure}

The following test consisted in exposing both technologies to two consecutive light pulses, which were generated by two LEDs. The pulses were separated by \qty{\sim 500}{ns}, and their intensity was initially set such that $\mathcal{O}$(\qty{1}{ photon}) was observed by the device for each pulse. Afterwards, the intensity of the second LED was increased progressively in consecutive measurements until the device response saturated. In the case of the analogue SiPM, at each step the readout scope resolution was adjusted so that both pulses were completely captured. The starting situation, with the two LEDs pulsated at low intensity, and the final situation, where one LED was pulsated at low intensity and the other was saturating the device, are presented in Fig.~\ref{fig:adcres} for the analogue and the digital cases. 

% \begin{figure}[tbp!]
% \centering

% % ---------- LEFT ----------
% \begin{minipage}[t]{0.49\textwidth}
% \centering
% \textbf{SiPMs}\par\vspace{0.5em}

% \includegraphics[width=\textwidth]{text/images/SiPMADCRes.pdf}
% \end{minipage}
% \hfill
% % ---------- RIGHT ----------
% \begin{minipage}[t]{0.49\textwidth}
% \centering
% \textbf{PDCs}\par\vspace{0.5em}

% \includegraphics[width=\textwidth]{text/images/PDCsADCRes.pdf}
% \end{minipage}

% \caption{SiPM response when exposed to two consecutive light pulses (left). The red line shows the case of two low-intensity pulses ($\mathcal{O}$(\qty{1}{photon})) with the oscilloscope resolution tuned to optimally capture this signal. The second light pulse is increased until the device is saturated (blue), and the oscilloscope resolution is optimised to capture both pulses. A degradation of the first pulse can be observed. Finally, the same pulse is captured, setting the oscilloscope resolution to the same value used in the original low-intensity configuration, thereby losing information about the second saturating peak (dashed black). In contrast, when the PDCs (right) are exposed to two low-intensity pulses (in dashed red), and then to a low-intensity pulse and a saturating pulse (solid blue), the first signal is observed without distortion in both cases. For both devices, the signals displayed are the averaged of 100 events.}
% \label{fig:adcres}

% \end{figure}

The left plot shows the analogue SiPM response, where it can be observed that lowering the resolution to resolve the large signal results in a loss of resolution for the small peak (blue line). However, if the resolution of the small peak is optimised (dashed line), information about the large pulse is lost due to ADC saturation. In other words, in analogue devices, a trade-off between single-photon resolution and dynamic range is required, which can become a complication in systems with large number of SiPMs. %A similar situation arises from the dynamic range of the amplification chain: achieving single-photon sensitivity requires significant amplification of the analogue SiPM signal, whereas large signals can saturate the amplifier, leading to information loss.

In the PDCs, both low- and high-intensity signals can be observed within the same acquisition without any change in the device’s parameter configuration. This is, again, a result of the discrete nature of the readout, where the signal is expressed in terms of SPAD activation per 10~ns bin. While the response remains subject to saturation at high occupancy, the signal representation is consistent across the entire dynamic range of the device.

%Finally, the possibility of disabling or enabling individual SPADs in the PDCs was explored. Fig.~\ref{fig:tcr} presents the measured DCR per individual SPAD for the four PDCs in a board. 87\% of the SPADs present less than \qty{\sim 300}{cps}. However, a small fraction of SPADs exhibit significantly higher rates, up to two orders of magnitude larger. These so-called \textit{screamer} SPADs can contribute disproportionately to the overall DCR, and are present in both digital and analogue SiPMs. However, while they cannot be selectively mitigated in analogue implementations, the PDC approach allows these channels to be masked, reducing their impact on the overall system behaviour.

Finally, the possibility of disabling or enabling individual SPADs in the PDCs was explored. The DCR per SPAD was evaluated, and it was found that a small fraction of them ($\sim$4\%) exhibited significantly higher rates, up to two orders of magnitude larger that the rest. These so-called \textit{screamer} SPADs can contribute disproportionately to the overall DCR, and are present in both digital and analogue~\cite{bib:screamers}. However, while they cannot be selectively mitigated in analogue implementations, the PDC approach allows these channels to be masked, reducing their impact on the overall system behaviour.

As an example, the same four PDCs were configured to store events whenever at least one SPAD from three different PDCs was activated within a \qty{20}{ns} coincidence window. First, the measurement was performed with all SPADs active, recording a rate of \qty{0.509 +-0.015}{coincidences / s}. The same measurement was then repeated after disabling all SPADs with a DCR above \qty{1}{kcps}, observing a rate of \qty{0.041 +-0.004}{coincidences/s}. This accounts for a reduction of the measured random coincidence rate by one order of magnitude by masking only \qty{4}{\%} of the SPADs. No further optimisation of the coincidence configuration was performed, as the goal was to demonstrate the masking capability.

%\begin{figure}[tbp]
%\centering
%\includegraphics[width=0.49\textwidth]{text/images/PDCsTCR.pdf}
%\caption{Dark count rate per SPAD in counts per second (cps). Different colours represent the 4 PDCs on the board. The red dashed line depicts the chosen threshold for considering a SPADS as a screamer (\qty{\sim e3}{cps}).
%\label{fig:tcr}}
%\end{figure}
\subsection{Exploring calorimetry potential}

%The PDCs were employed to characterise a radioactive spectrum. The spectrum was provided by a \ce{^{90}Sr} beta source coupled to a \qtyproduct{15x15x5}{\mm} \textit{EJ-228} plastic scintillator from \textit{Eljen}~\cite{bib:eljen}, painted with reflective paint to enhance light collection and coupled to the sensors with optical grease for light collection (see left of Fig.~\ref{fig:betas}).

\begin{figure}[tbp]
\centering
\begin{subfigure}{.45\textwidth}

\includegraphics[height=0.75\textwidth]{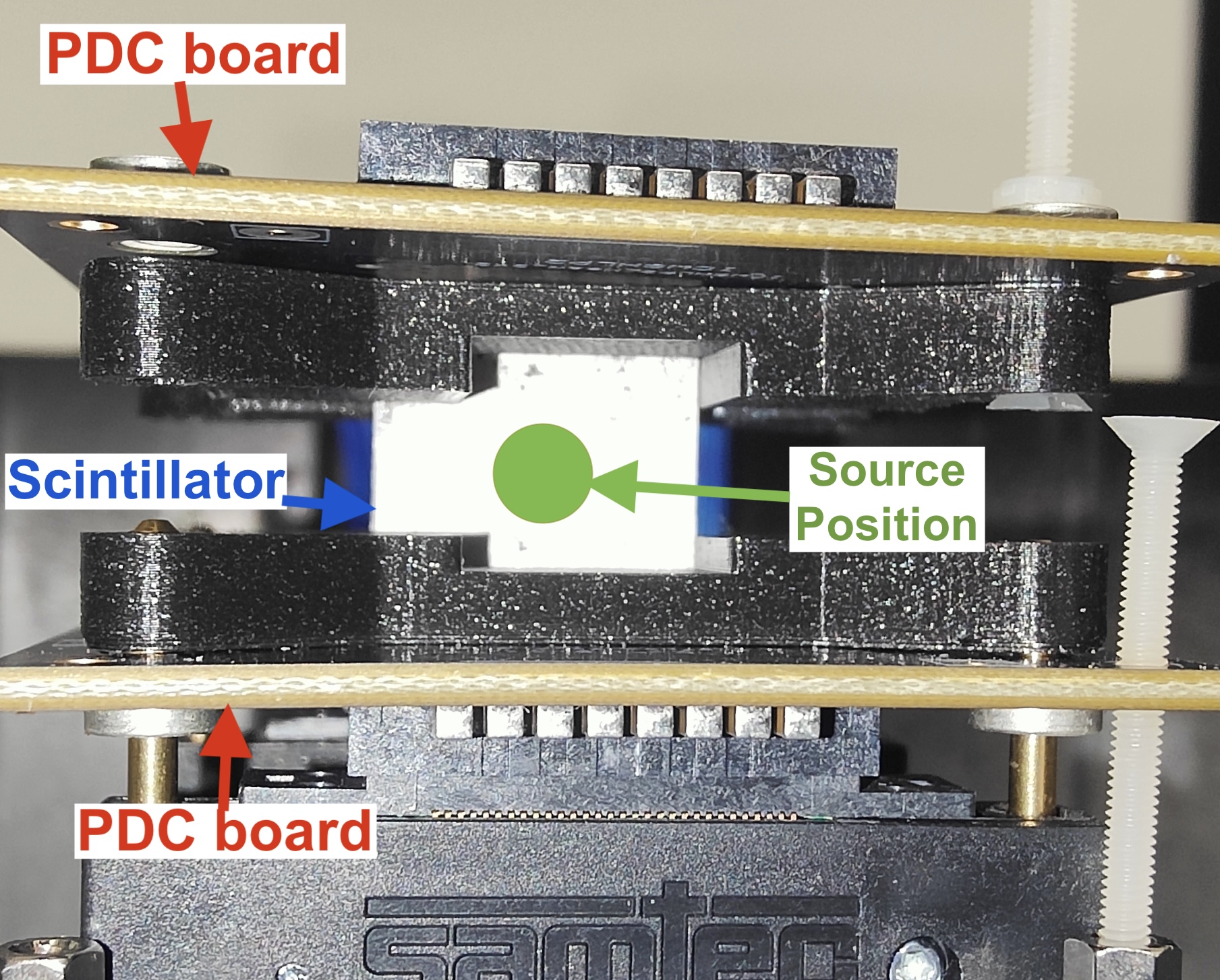}
\end{subfigure}
  \qquad
\begin{subfigure}{.45\textwidth}
\includegraphics[height=0.8\textwidth]{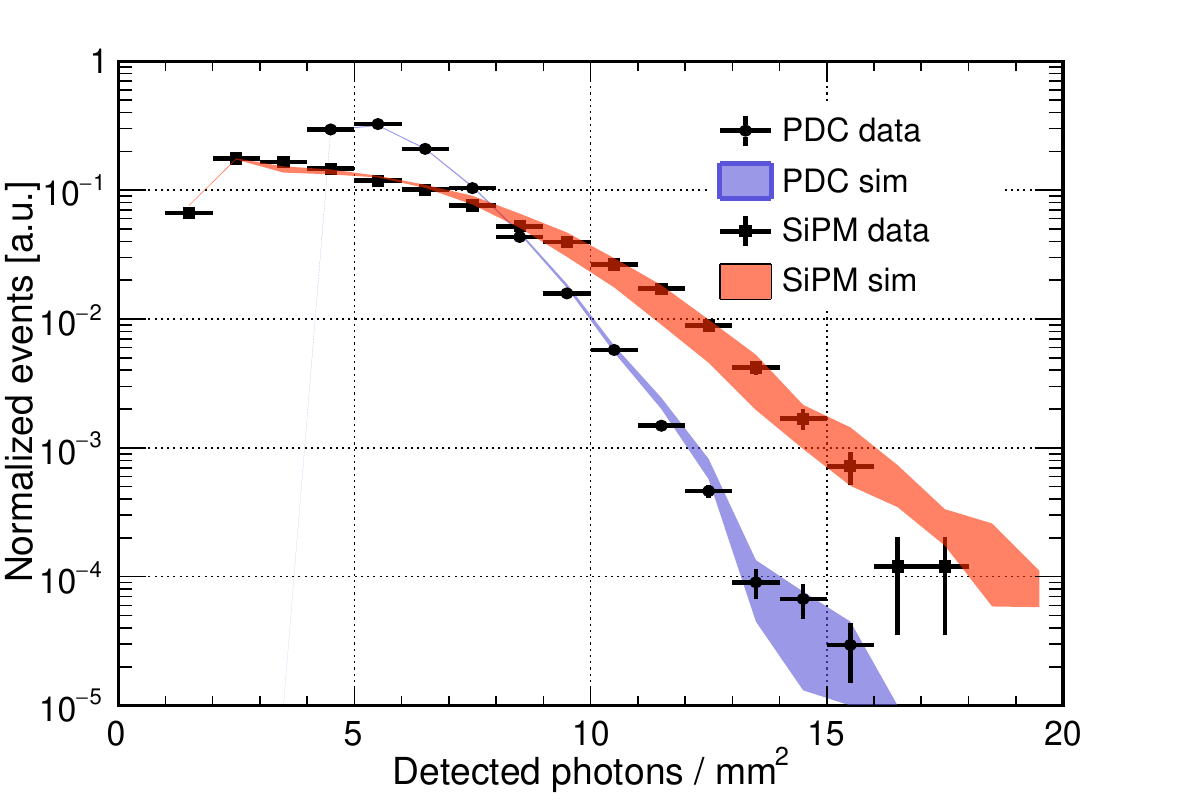}
\end{subfigure}

\caption{Left: Experimental setup for the beta spectrum measurement with the PDCs. Two sets of PDCs are placed on top and below a scintillator seen in white in the middle. The \ce{^{90}Sr} source is then inserted in the gap shown by the green dot. Note that for the SiPM case, a single SiPM was placed at the bottom. Right: Measured beta spectrum of \ce{^{90}Sr} using the SiPM (open circles) and the PDCs (solid circles), in comparison with the corresponding simulation (orange and lilac bands for analogue and digital, respectively), with a 1$\sigma$ shaded region. The simulation uncertainty arises from statistics and variations within the measured tolerance of the geometry of the setup. 
\label{fig:betas}}
\end{figure}

The PDCs were employed to characterise a radioactive spectrum. Measurements were first performed using a conventional analogue SiPM and subsequently repeated with the PDCs, using the same DAQ systems explained in the previous section in each case. In order to compensate for the significant optical coverage difference between technologies, a single SiPM was used compared to eight PDCs (grouped on two PCBs). The spectrum was provided by a \ce{^{90}Sr} beta source coupled to a \qtyproduct{15x15x5}{\mm} \textit{EJ-228} plastic scintillator from \textit{Eljen}~\cite{bib:eljen}, painted with reflective paint to enhance light collection. Both types of sensors were optically coupled with grease to the scintillator to optimize light transmission. A picture of the experimental setup with the PDCs can be seen on the left of Fig.~\ref{fig:betas}.

%Measurements were first performed using a conventional analogue SiPM and subsequently repeated with the PDCs, using the same DAQ systems explained in the previous section in each case. Both types of sensors were optically coupled with grease to the scintillator to optimize optical transmission. In order to compensate for the significant optical coverage difference between technologies, two PCBs (eight PDCs it total) were simultaneously used in this case (see Fig,~\ref{fig:betas}-left). 
In both cases, the observed spectra were compared with a Geant4 simulation to verify the overall consistency of the spectral shape and event rates at a qualitative level. Again, the goal of this section is not a quantitative performance comparison between the two technologies, but rather a comparison of the respective operational and analysis workflows by each readout approach under identical experimental conditions. 

For the analogue SiPM, the analysis followed a standard waveform-based reconstruction chain~\cite{bib:gundacker,bib:piemonte}. First, the response of the device and the readout electronics was calibrated to establish the relation between the observed signal and the number of detected photons, including corrections for known effects such as undershoot in the amplification chain. Trigger conditions were then defined to separate random dark counts from genuine scintillation signals, with an operating threshold on waveform height corresponding to approximately 10 detected photons, which was found to be appropriate for this configuration. Data were acquired in multiple runs with and without the beta source to account for additional background contributions, including cosmic-ray interactions and residual dark count activity. Each recorded waveform was subsequently processed to extract the corresponding photon-equivalent signal. Finally, the different datasets were combined, and the background contribution was estimated and subtracted from the source runs. The resulting spectrum is shown in Fig.~\ref{fig:betas}.

In the case of the PDCs, the analysis chain, although similar, is based on direct time-resolved readout of the activated SPADs. Since the output is intrinsically digital in the number of photons, no waveform integration or analogue charge reconstruction is required. Although calibration is not needed, other aspects must be evaluated before data taking, such as the trigger setup. The trigger configuration required at least one photon to be detected in at least five of the eight PDCs within a \qty{20}{ns} coincidence window, and this was set such that the background events represented less than \qty{1}{\%} of the signal events. As a result, the data processing pipeline is reduced to event selection based on coincidence conditions and the combination of the detected activity. Mimicking the analogue case, two runs (with and without the beta source) were taken, and the resulting spectrum, depicted in Fig.~\ref{fig:betas}, with background subtracted.

\begin{figure}[tbp]
\vspace{-2cm}
\centering
\includegraphics[height=0.95\textheight]{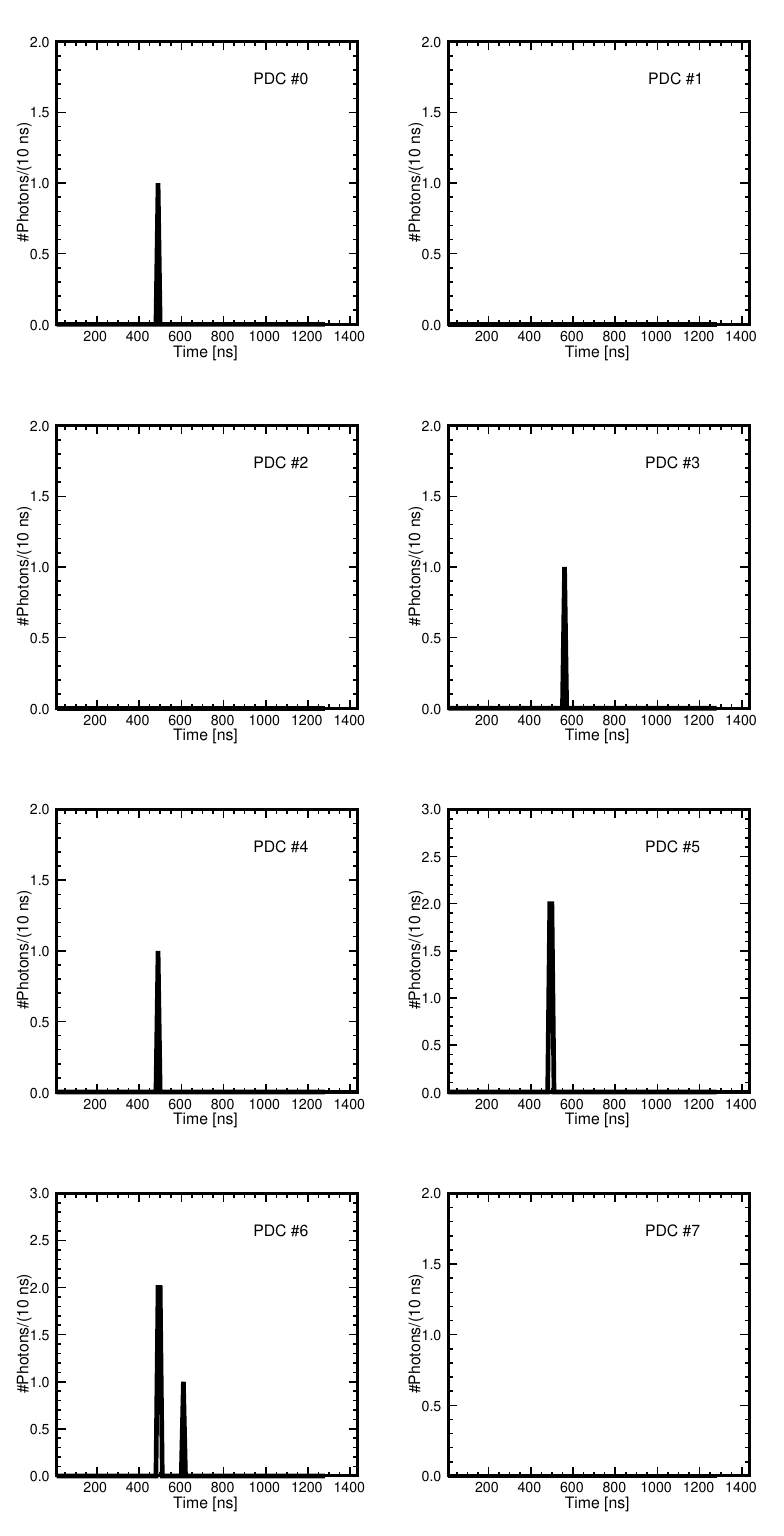}
\caption{Triggered event using the digital devices. Each plot represents an individual PDC waveform. PDCs \qtyrange{0}{3} and \qtyrange{4}{7} correspond each to their respective boards. A sharp pulse is observed in five of the eight chips at \qty{\sim 500}{ns}, with a potential dark count in PDC 6 at \qty{\sim 600}{ns}.
\label{fig:beta_event}}
\end{figure}

Fig.~\ref{fig:beta_event} shows a single event recorded by the PDCs as an illustrative example. It is worth noting that the event-related photons are concentrated in a very narrow time window around \qty{\sim 500}{ns}, suggesting that the second peak in PDC 6 is very likely a random dark count that falls within the acquisition window. This timing precision enables the implementation of additional noise-reducing strategies.

\subsection{Cosmic-ray identification}
%As a last exercise, PDCs were used to identify the passage of cosmic rays. For this application, two PDC boards with four chips each were stacked vertically (see Fig.~\ref{fig:cosmic_setup})-left, and each chip was optically coupled to a plastic scintillator as the one used in the previous section. Under this configuration, four different kind of coincidences were considered as presented in Fig.~\ref{fig:cosmic_setup}-right. In this particular case, the PDCs were configured to trigger when at least two chips observed at least three photons. Fig.~\ref{fig:cosmic_event} shows a triggered event where, again, it can be observed how clear PDCs' signals are. Finally, Fig.~\ref{fig:cosmics} shows the observed percentage of each coincidence type, in comparison with a simple G4 simulation.

\begin{figure}[tbp]
\centering
\includegraphics[height=0.3\textheight]{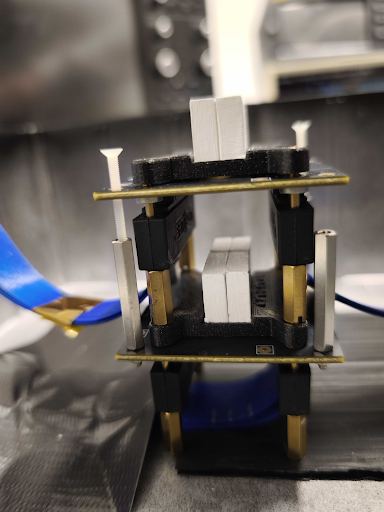}
\includegraphics[height=0.3\textheight]{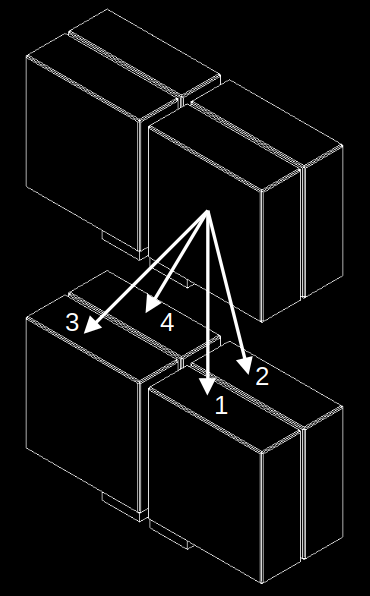}
\caption{Experimental setup for the cosmic-ray tracking configuration (left). The two PCBs are arranged in parallel, with a scintillator optically coupled to each PDC. The right panel shows the four types of coincidences considered for the study. %Taking the top-south piece as a reference for the example: 1) completely vertical, 2) long-side diagonal, 3) short-side diagonal and 4) corner diagonal.
\label{fig:cosmic_setup}}
\end{figure}

For the final use case, the PDC system was operated in a configuration designed to study coincident signals associated with cosmic-ray events. Having gained experience with the devices through the previous measurements, no comparison with conventional SiPMs was performed, and the system was operated independently, providing a first demonstration of standalone PDC operation in a particle-physics use case. Two 2$\times$2 PDC arrays were stacked vertically as shown in Fig.~\ref{fig:cosmic_setup}-left. Each chip was optically coupled to a plastic scintillator identical to that used in the previous section. Within this setup, we studied coincidence patterns between different channels as defined in Fig.~\ref{fig:cosmic_setup}-right. %Taking, for example, the top-bottom scintillator, the four types of coincidences would be: 1) completely vertical, 2 long-side diagonal, 3) short-side diagonal and 4) corner diagonal.

The analysis focuses on simple multiplicity-based trigger topologies across the stacked modules, which can be used to select candidate through-going events. In this configuration, the system was programmed to trigger when at least two PDCs registered at least three photon detections within a \qty{20}{ns} coincidence window.

%Fig.~\ref{fig:cosmic_event} shows an example of a triggered event. The recorded response illustrates the time-correlated activity across multiple channels, consistent with the expected behaviour of a traversing particle crossing the stacked scintillators.

\begin{figure}[tbp]
\centering
\includegraphics[width=0.49\textwidth]{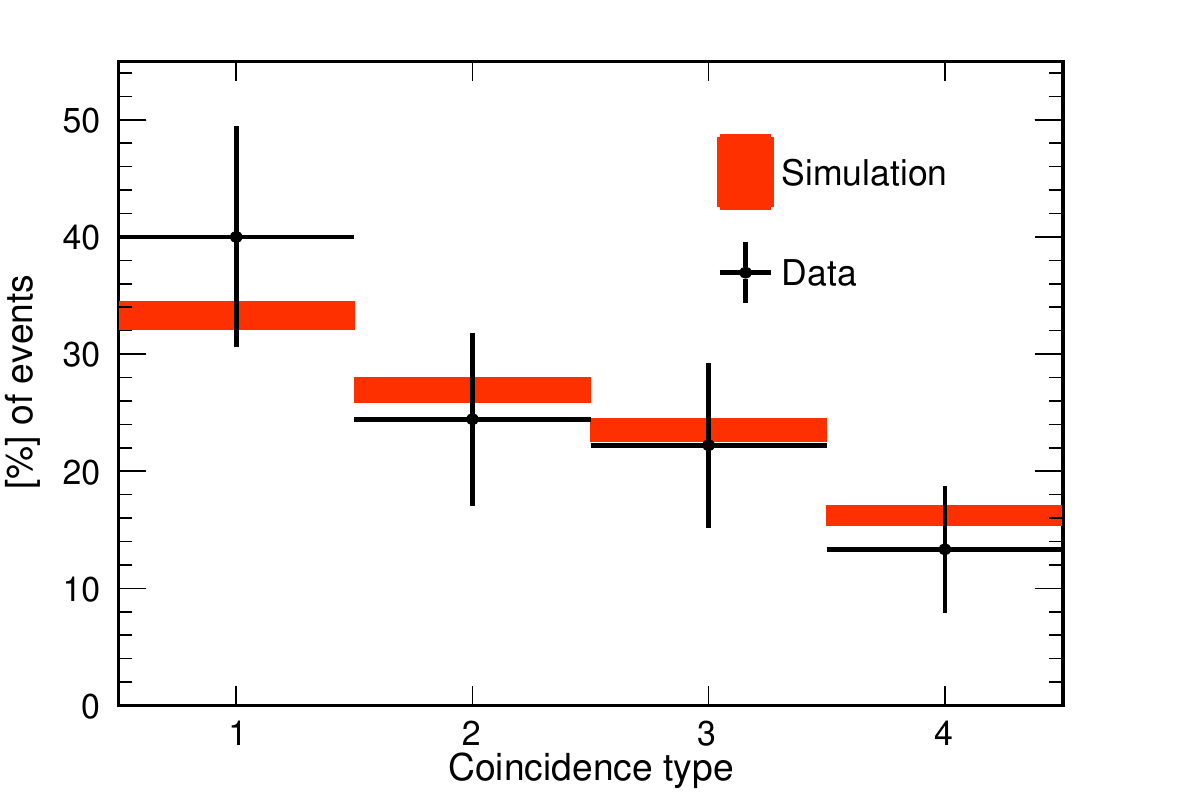}
\caption{Percentage of coincidence types, as depicted in Fig.~\ref{fig:cosmic_setup}-right, for data (points) and simulation (red areas) showing the 1$\sigma$ uncertainty for each, error bar for data and band for simulation.
\label{fig:cosmics}}
\end{figure}

Fig.~\ref{fig:cosmics} shows the measured distribution of the different coincidence classes, compared to a basic Geant4 simulation. As it can be observed, data and simulation are in good agreement, indicating that the devices are working as intended. %\sout{The comparison is intended as a qualitative cross-check of the observed event topology rather than a detailed validation of the detector response model.}

%\begin{figure}[htbp]
%\centering
%\includegraphics[height=0.95\textheight]{text/images/pdcs_cosmic_single_event.pdf}
%\caption{cosmic event?
%\label{fig:cosmic_event}}
%\end{figure}

\section{Conclusions and User Perspective}
%We have presented the first evaluation of PDCs in particle physics applications. First they have been exposed to a controlled light source, and afterwards used to characterize a beta spectrum and to identify the passage of cosmic rays. This has allowed us to observe their advantages with respect to their traditional analogue counterpart. 

%As their response remains digital through the whole readout chain, PDCs provide single-photon resolution across its whole dynamic range, avoiding the use of amplification chains and ADC-resolution limitations. The single photon response does not need to be calibrated nor disentangled from amplification chain issues, such as undershoot. This ultimately reduces the ambiguity when converting the output signal into number of photons detected, and when identifying consecutive light pulses with different amplitudes. Here we are not stating that these limitations cannot be overcome by analogue SiPMs but that digital SiPMs overcome them naturally. These advantages, alongside the possibility of deactivating high-DCR SPADs, make the overall physics studies presented in this work simpler, with the main limitation being learning to program the FPGA to operate the devices. 

%Even though these basic applications have been developed using prototypes of the digital devices, their benefits can be easily extrapolated to high energy physics experiments using photon detectors. The development of the fully integrated 3D devices is on its way. 

We present a simple demonstration of PDC operation in two particle-physics test cases. The prototype devices were first characterised using LED light sources and subsequently tested in two representative scenarios: measuring a beta spectrum with a plastic scintillator and identifying coincidence-based signatures associated with cosmic-ray events. These studies aim to understand PDC operation under simple cases and to anticipate their potential under realistic experimental conditions. Additionally, the agreement with the basic Geant4 simulations supports a correct understanding of the performance results presented.

Across all measurements, the PDC architecture provides a direct digital representation of SPAD activity, enabling time-resolved photon counting without requiring waveform integration of analogue signals. This avoids the need for analogue charge reconstruction and associated calibration of the amplification chain. As a result, the interpretation of the signal is based on discrete cell activation information rather than on integrated analogue amplitudes, simplifying the analysis chain in the user cases considered. Moreover, the direct time-resolved photon counting enables the temporal structure of the optical signals to be reconstructed over a wide range of timescales, from nanosecond to microsecond. This capability is particularly relevant for scintillation-based detectors, where the temporal profile of the light emission can extend over multiple timescales, as is the case for noble-element detectors. It should be emphasised that these observations do not imply conventional SiPM systems are incapable of comparable performance; rather, the digital architecture naturally provides a more direct representation of the detected signal.
%It should be emphasised that these observations do not imply that analogous performance cannot be achieved with conventional SiPM systems, but rather that the digital architecture naturally provides a more direct representation of the detected signal.

In addition, the ability to selectively disable individual high-noise SPADs offers a flexible handle to mitigate the impact of non-uniform pixel behaviour on coincidence-based measurements. This feature was shown to be effective in reducing accidental coincidence rates under the tested conditions, highlighting its potential relevance for low-threshold triggering applications. 

Although the devices used in this work are proof-of-concept prototypes with very low coverage, the results indicate that PDC architecture is a promising approach for future photon-detection systems in particle physics. Their potential advantages will ultimately depend on the specific integration strategy, readout implementation, and application requirements. Fully integrated 3D PDC implementations, that were not available at the time of this work, will further determine their applicability in large-scale detector systems. Of particular importance will be achieving the sub-\qty{100}{ps} time resolution to enhance time-of-flight determination capabilities, thereby positioning PDCs as viable alternatives for precision timing applications~\cite{bib:s21020598, bib:s23073376}.

From a user perspective, the main challenge encountered during this study was the initial learning curve associated with the digital architecture of PDCs. Configuring the devices through low-level registers, in particular, represents an additional barrier for users accustomed to conventional SiPMs or PMTs. Modern software-assistance tools, including large language models, can further facilitate the interpretation of device documentation and the development of the required configuration and control code, reducing the initial barrier to becoming familiar with the technology. Once this architecture is understood, however, the operation and subsequent data analysis are remarkably straightforward: the detector provides directly digitised photon counts, avoiding much of the signal processing typically required for analogue readout. The digital nature of the output also brings the acquired data closer to the quantities of direct interest for the physics analysis, considerably shortening the processing chain. The direct access to the temporal distribution of detected photons, currently provided in 10 ns time bins and with a future target of the order of 100 ps, offers an attractive and flexible approach to timing and coincidence measurements without relying on the analysis of extended analogue pulse shapes.  Overall, our experience suggests that, while PDCs require some initial familiarity with their digital architecture, their straightforward data handling and fine-grained temporal information make them an attractive technology for future particle-physics applications.

%oi meit, guay ar llu jier
\section*{Acknowledgements}

The authors acknowledge the support from the following agencies and institutions: the European Research Council (ERC) under Grant Agreement No. 951281-BOLD, and the Arthur B. McDonald Canadian Astroparticle Physics Research Institute (Queen’s University) towards the dissemination of the PDC technology within the astroparticle physics community. The authors also acknowledge the financial contributions to the work behind the PDC and tile controller from the Natural Sciences and Engineering Research Council of Canada (NSERC), the Fonds de recherche du Québec – Nature et technologies (FRQNT), the Regroupement stratégique en microsystèmes du Québec (ReSMiQ), and the Canada Foundation for Innovation (CFI). The design of the PDC and tile controller mentioned in this work was also supported in part by the United States Department of Energy, Office of Defense Nuclear Nonproliferation Research and Development in the National Nuclear Security Administration.

The authors would also like to acknowledge the support of DRD2, whose efforts in bringing the involved groups together and fostering this collaboration have been instrumental in enabling this work.

\bibliographystyle{elsarticle-num} 
\bibliography{bib}
%\bibliography{bib_short}



%% For example:

%% \bibitem[Aladro et al.(2015)]{Aladro15} Aladro, R., Martín, S., Riquelme, D., et al. 2015, \aas, 579, A101

%%\end{thebibliography}

\end{document}